\documentclass[amsmath,amssymb]{interact}

\usepackage{epstopdf}
\usepackage{xcccom}
\usepackage{mathrsfs}
\usepackage{mathtools}
\usepackage{tabularx}
\usepackage{threeparttable}
\usepackage{dcolumn}
\newcolumntype{d}[1]{D{.}{.}{#1}}
\usepackage[caption=false]{subfig}
\usepackage[numbers,sort&compress,merge]{natbib}
\usepackage{hyphenat}
\theoremstyle{plain}
\usepackage{braket}

\theoremstyle{definition}

\usepackage{multirow,dcolumn}
\theoremstyle{remark}

\newcommand{\Ts}{\mathscr{T}}
\begin{document}

\title{Spin-orbit coupling matrix elements
from the  explicitly connected expressions of the response functions within the coupled-cluster theory}

\author{
\name{A.~M. Tucholska\textsuperscript{a}\thanks{CONTACT A.~M.Tucholska Author. Email: am.tucholska@uw.edu.pl}, M. Lesiuk\textsuperscript{a}\thanks{Decicated to the memory of Professor Lutosław Wolniewicz}, R. Moszynski\textsuperscript{a}}
\affil{\textsuperscript{a} Faculty of Chemistry, University of Warsaw, Pasteura 
  1, 02-093 Warsaw, Poland}
}

\maketitle

\begin{abstract}
  In this work we present a coupled cluster based approach to the computation of the spin orbit coupling matrix elements.
  The working expressions are derived from the quadratic response function with the coupled cluster parametrization, using the auxiliary excitation
  operator $S$. Systematic approximations are proposed with the CCSD and CC3 levels of theory.
  The new method is tested by computing lifetimes of several electronic states of Ca, Sr and Ba atoms, with Gaussian and Slater basis sets. The results are compared with available theoretical and experimental data. 
\end{abstract}


\section{Introduction}

The spin-orbit (SO) interaction is a relativistic effect which plays a crucial role in the description of the electronic structure
of heavy atoms. It causes the fine structure splitting and mixing of  states with different multiplicities.
 In heavy atoms, the energy separation of the multiplet
states due to the SO interaction becomes comparable to the energy separation between different electronic states.
The second effect on the electronic structure is the mixing of states with  different multiplicities causing
radiative (phosphorescence)  and nonradiative (intersystem crossing) decays. Contributions of  spin-forbidden transitions are
crucial for the computation of lifetimes of electronic states.\cite{marian2001spin}

In this work, we focus on the spin-orbit coupling operators
introduced in the framework of  effective core potentials (SO ECP). This approximation was first proposed by Pitzer\cite{pitzer1997ab} and Schwartz,\cite{hafner1978pseudo}
and is used in this work in the Pitzer and Winter formulation.\cite{pitzer1988electronic} The effective Hamiltonian is given by
\equl{H_{\rm{SO}} = \sum_{l=1}^{l_{\mt{max}}}\xi_l(r)P_l\vec{l}\cdot\vec{s}P_l
}{spinorb}
where $\xi_l(r)$ is a radial potential and $P_l = \sum\limits_{m_l}\ket{l,m_l}\bra{l, m_l}$
is the projection operator onto spherical harmonics with angular
momentum $l$ placed at a given pseudopotential center.

In high-resolution spectroscopy, interpretation of the experimental spectra requires a theory that
effectively includes both the relativistic effects and the electron correlation at a high level.
In this work we propose  computing the SO coupling matrix elements within the
coupled cluster (CC) approach to the response theory. 
There exist two approaches to the coupled cluster response theory. The first one, introduced
in 1977 by Monkhorst\cite{monkhorst1977calculation, dalgaard1983some} and extended by
Bartlett et. al, \cite{jorgensen2012geometrical, fitzgerald1986analytic, salter1987property, salter1989analytic}
is based on  differentiation of the CC energy expression. Later, Koch et al. \cite{koch1990coupled, koch1994caclculation,
  koch1997coupled, christiansen1998response} derived the
time-averaged quasi-energy Lagrangian technique. Within this approach,
the expressions for the linear, quadratic, and cubic
response functions,\cite{christiansen1998response}
transition moments,\cite{christiansen1998integral} spin-orbit coupling
matrix elements,\cite{christiansen2002radiative, helmich2016spin} and
many other properties,\cite{helgaker2012recent}
at the CC2,\cite{christiansen1995second, hald2000linear} CCSD,\cite{koch1994caclculation}
 and CC3 levels \cite{koch1997cc3, hald2002calculation} of theory were proposed.

In this work we use the  an alternative approach in which one computes molecular properties directly from the
average value of an operator $X$, using the auxiliary excitation operator $S$. 
Parenthetically we note that this method, referred to as XCC,
should not be confused with the approach of Bartlett and Noga\cite{bartlett1988expectation} of the same name.
The  operator $S$ was introduced in the context of CC theory in the works of Arponen et al.\cite{arponen1983variational, arponen1987extended, bishop1990correlations} The authors defined the operator $S$ through a set of nonlinear equations with no
systematic scheme of approximations. In 1993 Jeziorski and Moszynski \cite{jeziorski1993explicitly} proposed a different expression for
the operator $S$ that satisfies a set of  linear equations and can be systematically approximated. Starting from the formula
for the average value of $X$  expressed through a finite series of commutators, the XCC method was extended to the computation of various 
electronic properties: electrostatic\cite{moszynski1993many} and exchange\cite{moszynski1994many}
contributions to the interaction energies of closed-shell systems,
the polarization propagator, \cite{moszynski2005time} first-order molecular properties,\cite{korona2006one} static and dynamic dipole polarizabilities, \cite{korona2006time} and transition moments between
excited states.\cite{tucholska2014transition, tucholska2017transition}
There is also a series of publications by Korona et al. on the computation of
symmetry-adapted perturbation theory (SAPT) contributions.\cite{korona2008first, korona2008second, korona2008dispersion, korona2009exchange}
An extensive review of the XCC methodology was recently published.\cite{tucholska2021molecular}
\section{XCC theory}
In  CC theory \cite{coester1958bound, coester1960short, vcivzek1966correlation, cizek1969use} the ground state wave function $\Psi_0$  is represented by the
exponential ansatz $\Psi_0 = e^T\Phi$, where $\Phi$ is the Slater determinant
and
the cluster operator $T$ is a sum of $n$-tuple excitation operators
\equ{
  T = \sum_{n=1}^N T_n
}
where $N$ is the number of electrons.
Each of the cluster operators $T_n$ is represented by a product of spin-free excitation operators $E_{ai}$\cite{paldus1988clifford}
\equ{T_n = \frac{1}{n!}\sum_{\mu_n}^{N} t_{\mu_n}\mu_n = \frac{1}{n!}\sum_{\mu_n}^{N} t_{\mu_n}E_{ai}E_{bj}\ldots,
}
where $\mu_n$ denotes $n$-th excitation level.
The indices $a, b, c \ldots$, $i, j, k \ldots$ and $p, q, r \ldots$ denote the virtual, occupied, and general orbitals, respectively.

 In the XCC method molecular properties are computed with the use of the auxiliary excitation operator $S= S_1 + S_2 + S_3 + \ldots$ defined as
 \equl{e^S\Phi = \frac{\etd \et \Phi}{\braket{\etd\et}}.
 }{r3}
 The $S$ operator satisfies a set of linear equations expressed through a finite series of commutators
 \cite{jeziorski1993explicitly}
\equal{
S_n &= T_n  - \frac1n \hat{\mathscr{P}}_n \left ( \sum_{k=1}
\frac{1}{k!}\cm{\widetilde{T}\dgg}{T}_k \right ) \\
&- \frac1n \hat{\mathscr{P}}_n\left (\sum_{k=1}\sum_{m=0}
\frac{1}{k!}\frac{1}{m!} [\cm{\widetilde{S}}{\dg{T}}_k,T]_m\right)\nonumber,
}{ss}
where
 \equ{
\widetilde{T} = \sum_{n = 1}^{N} nT_n \qquad\widetilde{S} = \sum_{n = 1}^{N} nS_n,
}
and  $[A, B]_k$ denotes a $k$-tuply nested commutator.
 The superoperator $\hat{\mathscr{P}}_n(X)$ 
yields the 
excitation part of $X$
\equl{\hat{\mathscr{P}}_n(X) = \frac{1}{n!}\sum_{\mu_n} \braket{\mu_n|X}\mu_n}{superp}
and $\bra{\mu_n}$ and $\ket{\nu_m}$ constitute a biorthonormal basis
$\langle \mu_n|\nu_m\rangle = \delta_{\mu_n\nu_m}$.
 The expression for the transition moment is derived from the quadratic response function,
 $\braket{\braket{X;Y,Z}}_{\omega_Y, \omega_Z}$, which describes
the response of an observable $X$
to periodic perturbations $Y$ and $Z$ with frequencies $\omega_Y$ and $\omega_Z$, respectively.
The explicit form of the response function written as a sum over states reads\cite{christiansen1998response}
\equal{\langle\langle X; Y, Z \rangle \rangle&_{\omega_Y, \omega_Z}
=P_{XYZ}
\sum_{\substack{K'=1\\N'=1}}\quadrup{Y}{X}{Z},
}{quadrubig}
where the operator $P_{XYZ}$ symmetrizes the expression  with respect to    exchange of the indices 
 $X$, $Y$, and $Z$.
The indices $K'$ and $N'$ run over all possible excited states with excitation energies $\omega_{K'}$ and $\omega_{N'}$.
The transition moment $\TT_{KN}^X$ between the excited states $K$ and $N$, with $K \ne N$, is  computed as a residue of the quadratic response function 
\equal{&\lim_{\omega_Y\rightarrow -\omega_K}(\omega_K + \omega_Y)\lim_{\omega_Z\rightarrow \omega_N}(\omega_N - \omega_Z) \odpq\\
   &= \braket{\Psi_0|Y|\Psi_K}\braket{\Psi_K|X_0|\Psi_N}\braket{\Psi_N|Z|\Psi_0} =\TT^Y_{0K}\TT^X_{KN}\TT^Z_{N0}\nonumber.
}{tmexact}
where $\xmz = X - \langle X \rangle$ and $\langle X \rangle = \langle \Psi_0 | X | \Psi_0 \rangle$. 

 In our previous work \cite{tucholska2017transition} we employed the
 XCC formalism to express the quadratic response function in terms of the $T$ and $S$ operators. 
By using the first-order perturbed wave function  
\equ{\Psi^{(1)}(\omega_Y) = \mathcal{R}_{\omega_Y} Y|\Psi_0\rangle}
expressed through the resolvent $\mathcal{R}_{\omega_Y}$
 \equ{\mathcal{R}_{\omega_Y} = \sum_{N'=1} \frac{|N' \rangle \langle N'|}{\omega_N + {\omega_Y}},}
 \fr{quadrubig} is rewritten in the following form
\equl{ \quadraom = P_{XYZ}\langle \Psi^{(1)}(\omega_Y)|\xmz|\Psi^{(1)}(-\omega_Z) \rangle.
}{podst}
At this point we introduce the coupled cluster parametrization.
The normalized ground-state wave function in this parametrization is given by
\equl{|\Psi_0\rangle = \frac{|\et\Phi\rangle}{\langle \et\Phi|\et\Phi\rangle^{\frac12}},}{psiz}
and the first-order 
wave function $\Psi^{(1)}(\omega_Y)$ 
is defined as
\equl{|\Psi^{(1)}(\omega_Y)\rangle =  (\Omega_0^Y + \Omega^Y(\omega_Y)) |\Psi_0\rangle, }{psi1}
through the excitation operator
\equ{\Omega^Y(\omega_Y) = \Omega_1^Y(\omega_Y) + \Omega_2^Y(\omega_Y) + \ldots}
where the $n$-th excitation operator $\Omega_n^Y(\omega_Y)$
is 
\equl{
\Omega_n^Y(\omega_Y) = \sum_{\mu_n}\frac{1}{n!} O_{\mu_n}^Y(\omega_Y) \mu_n.
}{omg}
The operator $\Omega^Y(\omega_Y)$ is found 
  from the linear response equation
\cite{moszynski2005time, korona2006time} 
\equl{
  \braket{\mu_n|\cm{e^{-T} H e^T}{ \Omega^Y(\omega_Y)} + \omega_Y\Omega^Y(\omega_Y)
  + e^{-T}Ye^T} = 0.
}{om}
By transforming the $\mu_n$ index to the basis of the right, $r_N$, and left, $l_N$, eigenvectors  of the
CC Jacobian matrix\cite{sekino1984linear, koch1990coupled, helgaker2013molecular} $A_{\mu_n\nu_m} = \braketd{\mu_n}{\cm{e^{-T}He^T}{{\nu}_m}}$
 we arrive at
\equ{
\omega_NO_N^Y(\omega) +
\omega_Y O_N^Y(\omega)
  + \xi_N^Y= 0
}
where 
\equl{\xi_N^Y = \braket{l_N|e^{-T}Ye^T}.
}{xin}
With the explicit expression for $\Omega(\omega_V)$ given above,
\fr{podst} is reformulated using the following identities
\equa{&\Omega(\omega_Y) \et = \et \Omega(\omega_Y)\label{eql1},\\
  &\esdm\Phi = \Phi,\nonumber\\
  &X\Phi = \langle X\rangle\Phi + \phm(X)\Phi,\nonumber\\
  &\frac{\braket{\et|X|\et}}{\mian} = \sred{\esd\etm X \et \esdm}.\nonumber\label{eql2}
}
The quadratic response function becomes 
\equa{&\quadraom = \\
&= P_{XYZ}\sum_{\substack{K'=1\\N'=1}}\frac{\braket{\etm Y\et|l_{K'}}}{\omega_{K'} + \omega_Y}\frac{\braket{l_{N'}|\etm Z\et}}{\omega_Z - \omega_{N'}}\nonumber\\
&  \times\braket{\karkp|\esd\etm \xmz\et\esdm | \etrnp},\nonumber}
where we introduced a shorthand notation
\equs{
&\karkp = \phm\left(\esm\etd r_{K'}\etdm\es\right)\\
&\etrnp = \phm\left(\esd r_{N'}\esdm\right).
}
The transition moment between the excited states, $\TT_{KN}^X$ is obtained from the residue of the quadratic response function\cite{tucholska2017transition} by dividing the RHS of \fr{tmexact} by $|\TT_{0K}^Y\TT_{N0}^Z| = \sqrt{|\TT_{0K}^Y|^2|\TT_{N0}^Z|^2}$.
\equl{\TT_{KN}^X 
     = \frac{\braket{\kark|\esd\etm \xmz\et\esdm \etrn}}{\sqrt{\braket{\kark|\etrk}\braket{\karn|\etrn}}}.
}{glowne}
This expression was previously used to compute numerous dipole and quadrupole transitions between electronic states\cite{tucholska2017transition}.
In this  work we employed $\TT_{KN}^X $ to compute the spin-orbit transition moments.

\section{Computational details}
\subsection{Spin-orbit coupling matrix elements in the spin-free formulation}

The transition moments are usually
represented in the literature via the line strength $\mathcal{S}$, which is defined as the square of the
 transition moment from the state $K \equiv \bra{L'S'J'}$ to the state $N\equiv \ket{LSJ}$ \cite{shortley1940computation}
\equl{\mathcal{S}_{KN} = |\TT_{KN}^{\pmb{\Ts}^{(k)}}|^2=|\brakettred{ L'S'J'}{\pmb{\Ts}^{(k)}}{ LSJ}|^2. 
}{lsls}
In the above notation, the transition moment  $\brakettred{ L'S'J'}{\pmb{\Ts}^{(k)}}{ LSJ}$ is the
reduced matrix element of the irreducible tensor operator 
$\pmb{\Ts}^{(k)}=(\Ts_{-q}^{(k)}\ldots, \Ts_q^{(k)})$ of rank $k$ with $2k+1$ components $q$,
$q \in (-k,\ldots, k)$.   $L$, $S$, and $J$ are quantum numbers of the orbital, spin, and total angular momentum,
respectively.

It is a common practice in ab initio calculations, including our XCC implementation, to compute the transition moments in the point group symmetry basis $|{}^{2S+1}\Gamma^{m_S}\rangle$.
In order to compare our results with the available experimental and theoretical results we need to 
express the $\mathcal{S}_{KN}$ in the molecular point group symmetry basis.
This is done by transforming the $\ket{LSJ}$ basis to the  $| Lm_LSm_S\rangle$ using angular momentum algebra. The correspondence between  $| Lm_LSm_S\rangle$ and $|{}^{2S+1}\Gamma^{m_S}\rangle$ is straightforward.

 The reduced matrix element of \fr{lsls} can be written explicitly by summing over all  $m_J$ and $m_{J'}$ in the $| LSJm_J\rangle$ basis, 
 \equl{\mathcal{S}_{KN} = \sum_{m_J, m_{J'}}|\brakett{ L'S'J'm_{J'}}{\pmb{\Ts}^{(k)}}{ LSJm_{J}}|^2.
 }{lsls-nr}
 Instead of summing all these contributions, we exploit the Wigner-Eckart therem and express $\mathcal{S}_{KN}$ through
 the $3-j$ coefficients as
 \equl{\mathcal{S}_{KN} = |\left( (-1)^{(J'-m_{J'})} 
  \left( \begin{array}{ccc}
   J' & k &J\\
   -m_{J'} & q &m_J\\
  \end{array}  \right) \right)^{-1}\brakett{ L'S'J'm_{J'}}{\Ts_q^{(k)}}{ LSJm_{J}}|^2.
 }{ls-3j}
 It is important to note that the line strength $\mathcal{S}_{KN}$ does not depend on the choice of $m_J$, so
 any non-trivial contribution to the reduced matrix element of \fr{ls-3j} is sufficient to obtain the line strength.
Next, we  transform the
$| LSJm_J\rangle$ vetor to the $| Lm_LSm_S\rangle$ basis,
with the help
of the Clebsch\textendash Gordan coefficients
\equ{\ket{LSJm_J} = \sum_{m_L = -L}^{L} \sum_{m_S=-S}^S C^{Jm_J}_{Lm_LSm_S}\ket{Lm_LSm_S}.
}
Therefore the expression for the line strength becomes
\equa{\mathcal{S}_{KN} &= 
   |  (-1)^{(J'-m_{J'})} \left(\begin{array}{ccc}
   J' & k &J\\
   -m_{J'} & q &m_J\\
  \end{array}\right) ^{-1}\\
  &\times\sum_{\tiny{\substack{m_L=-L\\m_{L'}=-{L'}}}}^{L,L'}\sum_{\tiny{\substack{m_S=-S\\m_{S'}=-{S'}}}}^{S,S'}
  C^{Jm_J}_{Lm_LSm_S}C^{J'm_{J'}}_{L'm_{L'}S'm_{S'}}
  \brakett{ L'm_{L'}S'm_{S'}}{\Ts_q^{(k)}}{ \lsm}|^2.\nonumber
}

The effective spin-orbit operator $H_{\rm{SO}} $ from \fr{spinorb} is expressed through the triplet excitation operators
\equa{
  &T_{pq}^{11} = -a_{p\alpha}^\dagger a_{q\beta}\\
  &T_{pq}^{1-1} = a_{p\beta}^\dagger a_{q\alpha}\nonumber\\
  &T_{pq}^{10} = \frac{1}{\sqrt{2}} (a_{p\alpha}^\dagger a_{q\alpha}- a_{p\beta}^\dagger a_{q\beta})\nonumber,
}
resulting in
\equ{H_{\rm{SO}} =  \sum_{pq}\left[\left(-\frac{i}{2}V_{pq}^x -\frac{1}{2}V_{pq}^y\right)T_{pq}^{11}
  + \left(\frac{i}{2}V_{pq}^x-\frac{1}{2}V_{pq}^y\right)T_{pq}^{1-1}
  + \frac{i}{2\sqrt{2}}V_{pq}^zT_{pq}^{10}\right],
}
where $V_{pq}^v$
\equ{ V_{pq}^v =\frac{1}{i}\int \phi_p^\star(\mathbf{r})\xi(r)l_v P_l\phi_q(\mathbf{r})d\mathbf{r} \qquad v \in(x,y,z).
}
The transition moment $\TT_{KN}^X$, where $X=H_{\rm{SO}}$, 
from state $K\equiv\lsmp$ to $N\equiv\lsm$ becomes
\equa{&\brakett{\lsmp}{H_{\rm{SO}}}{\lsm}=\\
    &=\frac12\sum_{pq}\brakett{L'm_{L'}}{(iV_{pq}^x +V_{pq}^y)}{Lm_L}\brakett{S'm_{S'}}{T_{pq}^{10}}{Sm_S}\nonumber\\
   &+\frac12\sum_{pq}\brakett{L'm_{L'}}{(-iV_{pq}^x+V_{pq}^y)}{Lm_L}\brakett{S'm_{S'}}{T_{pq}^{10}}{Sm_S}\nonumber\\
   &+ \frac{i}{2\sqrt{2}}\sum_{pq}\brakett{L'm_{L'}}{V_{pq}^z}{Lm_L}\brakett{S'm_{S'}}{T_{pq}^{10}}{Sm_S}.\nonumber
}
To separate the spin and angular parts and use  the spin-free formalism, we expressed the $m_S$-changing spin-tensor operators $T_{pq}^{11}$ and $T_{pq}^{1-1}$ in terms of $T_{pq}^{10}$ by the
 virtue of the Wigner-Eckart theorem.
\subsection{Approximations}
To compute the XCC properties, one needs to follow four independent steps: obtain the amplitudes $T$ and $S$,
compute the excitation amplitudes $r_N$, and finally use \fr{glowne} to compute $\TT_{KN}^{X}$.
The calculation of the amplitudes $T$ is done by any standard CC method.
In this work we use the coupled cluster method
limited to single and double excitations (CCSD) and the coupled cluster method limited to
single, double, and approximate triple excitations (CC3).

The expression for the auxiliary amplitudes $S$, \fr{ss}, is a finite expansion, though 
it contains terms of high order in the fluctuation
potential,\cite{jeziorski1993explicitly} i.e., in the sense of many-body perturbation theory (MBPT).
To reduce the computational cost $S$ can be systematically approximated while retaining the size-consistency.\cite{tucholska2014transition}
Let $S_n(m)$ denote the $n$-electron part of $S$, where all contributions up to
and including the order $m$ of MBPT are accounted for. 
In the computations based on the CC3 model, we employ
\equsl{
S_1(2) &= T_1   \\
S_1(3) &= S_1\dr +  \hat{\mathscr{P}}_1\left ([T_1^{\dagger}, T_2] \right )  
+ \hat{\mathscr{P}}_1\left ([T_2^{\dagger}, T_3] \right )   \\
S_2(2) &= T_2  \\
S_2(3) &= S_2\dr + \frac12\hat{\mathscr{P}}_2\left ([[T_2^{\dagger}, T_2], T_2] \right )   \\
S_3(2) &= T_3
}
{s-approx}
where the CC3 equations for  $T_1$, $T_2$ and $T_3$ are given by Koch et al.\cite{koch1997cc3}
In the instances where the underlying model of the wave function is CCSD,
 we employ $S = S_1(3) + S_2(3)$ neglecting the terms containing $T_3$.
 The amplitudes $r_N$ are obtained from the EOM-CC3 or EOM-CCSD model depending on which approximation is used for the ground state, 
 with the cost scaling as
as $N^7$ and $N^6$, respectively.

The most challenging problem is to systematically approximate the transition moment formula.
In order to do so, we performed a commutator expansion of the numerator of \fr{glowne} and assessed the importace of each term individually by the order of its leading MBPT contribution
The formulas were derived automatically by the 
program \paldus~developed by one of us (AMT). Due to the computational and memory restrictions some
 additional approximations were used, as described below.

All terms resulting from commutator expansion of \fr{glowne} are of the type:
\equ{\braket{[[\mu_n,T^\dagger]_{k_1},S]_{k_2}|[[X,T]_{k_3},S^\dagger]_{k_4} | [\nu_m,S^\dagger]_{k_5}},
}
where $k_1-k_5$ are integers that denote the  nesting level and $m$ and $n$ are  the
excitation levels. 
For clarity we omit the  excitation levels of $T$ and $S$ in the above expression.
We include all  terms up to a given MBPT order with a few exceptions that are listed in \Frt{rzedy}. One should interpret the
description as:
\begin{itemize}
\item ``\emph{neglect $\langle\mu_n |\ldots|\nu_m\rangle$}'' means that all terms  that have $n$-tuple excitations in the
  bra and $m$-tuple excitations in the ket are neglected.\\
  \item ``\emph{neglect $\langle\mu_n |\ldots|\nu_m\rangle$ unless includes $T_1$ or $S_1$}''
    means that all terms  that have $n$-tuple excitations in
    the bra and $m$-tuple excitations in  the ket are neglected unless
    the operators $T_1$ or $S_1$ appear at least once, e.g., $\braket{X[\mu_3, T^{\dagger}_2] | [\nu_2, S^{\dagger}_1]}$
      is included, but $\braket{X[S_2, [\mu_3, T^{\dagger}_3]] | \nu_2}$ is not. 
      \\
    \end{itemize}
\begin{table}[!ht]
  \begin{center}
    \caption{Terms included in the XCC transition moments calculations.}\label{rzedy}
    \renewcommand{\arraystretch}{1.5}
  \begin{tabular}{cll}
         \hline
         leading MBPT order& CCSD& CC3\\
         \hline
                  0& all &all \\
                  \hline
                  1& all&all \\
                  \hline
                  2&all & neglect $\langle\mu_3 |\ldots|\nu_3\rangle$\\
                  \hline
                  \multirow{3}{*}{3}& \multirow{3}{*}{all} &neglect $\langle\mu_3 |\ldots|\nu_3\rangle$ \\[-5pt]
        &      &neglect $\langle\mu_2 |\ldots|\nu_3\rangle$ unless includes $T_1$ or $S_1$\\[-5pt]
        &      &neglect $\langle\mu_1 |\ldots|\nu_3\rangle$ unless includes $T_1$ or $S_1$\\
       \hline
  \end{tabular}
  \end{center}
\end{table}

\section{Numerical results}

\subsection{Spin-forbidden transitions}
The transition probability $\matha_{KN}$ from an initial state $K$ to a final state $N$ for the
dipole (E1) and quadrupole (E2) transitions is given by the Einstein coefficients
\equl{ \matha_{KN}(E1) = \frac{16 \pi^3}{3h\epsilon_0\lambda^3(2J_K+1)}\mathcal{S}_{KN}(E1),
}{e1}
\equl{ \matha_{KN}(E2) = \frac{16 \pi^5}{15h\epsilon_0\lambda^5(2J_K+1)}\mathcal{S}_{KN}(E2),
}{e2}
where $h$ is the Planck constant, $\epsilon_0$ is the vacuum permittivity, $\lambda$ is the wavelength in $[m]$, $J$ is the total angular momentum for the initial
state, $\maths_{KN}(E1)$ is the line strength of a dipole transition, and $\maths_{KN}(E2)$ is the line strength of a quadrupole transition.
Let us denote the electronic states $K$ and  $N$ from \fr{lsls} as $\ket{\psi_K}$ and $\ket{\psi_N}$, so that the line strength
is written down as
\equ{\maths_{KN} = |\TT_{KN}^{\pmb{\Ts}^{(k)}}|^2=|\braket{\psi_K||\pmb{\Ts}^{(k)}||\psi_N}|^2.
}
To derive the expression for the spin-forbidden transitions,
we use the Rayleigh–Schr\"{o}dinger perturbation theory (RSPT),\cite{oddershede1985calculation} where $H_{\rm{SO}}$ is treated as a perturbation. 
Assuming that we have an initial triplet state $\bra{^{3}\Psi_K}$ and a final singlet state $\ket{^{1}\Psi_N}$, the RSPT expansion
 of the ket wave function is given by
\equal{\ket{^3\Psi_K} & = \ket{^3\Psi_K^{(0)}} + \ket{^3\Psi_K^{(1)}} + \ldots \nonumber\\
  & =\ket{^{3}\psi_K^{(0)}} + \sum_{L} \frac{\braket{^{1}\psi_L^{(0)}|H_{\rm{SO}}|^{3}\psi_K^{(0)}}}
  {^{3}E_K^{(0)}-\prescript{1}{}E_L^{(0)}}\ket{^{1}\psi_L^{(0)}}+
  \sum_{L} \frac{\braket{^{3}\psi_L^{(0)}|H_{\rm{SO}}|^{3}\psi_K^{(0)}}}
  {^{3}E_K^{(0)}-\prescript{3}{}E_L^{(0)}}\ket{^{3}\psi_L^{(0)}} + \ldots
}{trip-rs}
where $\Psi$ denotes the spin-orbit coupled state, $\psi$ is a pure $LS$ state and  $^{m}E_L^{(0)}$ denotes the
zeroth-order energy of the $L$th state with multiplicity $m$.
The final ground state is
\equa{\ket{^1\Psi_N} & = \ket{^1\Psi_N^{(0)}} + \ket{^1\Psi_N^{(1)}} + \ldots \nonumber \\
  & =\ket{^{1}\psi_N^{(0)}} + \sum_{L} \frac{\braket{^{3}\psi_L^{(0)}|H_{\rm{SO}}|^{1}\psi_N^{(0)}}}{^{1}E_N^{(0)}-\prescript{3}{}{E}_L^{(0)}}\ket{^{3}\psi_L^{(0)}} + \ldots
}
$L$ runs only over states with  triplet multiplicity, as other terms vanish due to the selection rules.
Expanding the electric dipole perturbation we get
\equl{ \TT_{KN}^{\mathdm} = \braket{\Psi_K^{(0)} + \Psi_K^{(1)}|{\mathdm}|\Psi_N^{(0)}+\Psi_N^{(1)}},
}{tm-rs}
where we neglect higher-order terms.\cite{marian2012spin} 
Additionally, we set  $m=1$ in the 
expansion (\ref{trip-rs}), as states of other multiplicities are not directly  connected by the dipole transition with the ground state.
The term $\braket{\Psi_K^{(0)}|\mathdm|\Psi_N^{(0)}}$  vanishes due to the selection rules, so the final expression for the transition
dipole moment is given by
\equl{\TT_{KN}^{\mathdm}=
  \sum_{L} \frac{\braket{^{3}\psi_L^{(0)}|H_{\rm{SO}}|^{1}\psi_N^{(0)}}}{^{1}E_N^{(0)}-\prescript{3}{}E_L^{(0)}}\braket{^{3}\psi_K^{(0)}|\mathdm|^{3}\psi_L^{(0)}}
  + \sum_{L} \frac{\braket{^{1}\psi_L^{(0)}|H_{\rm{SO}}|^{3}\psi_K^{(0)}}}{^{3}E_K^{(0)}-\prescript{1}{}E_L^{(0)}}\braket{^{1}\psi_L^{(0)}|\mathdm|^{1}\psi_N^{(0)}}.
}{spin-forb}
The radiative lifetime\cite{drake2006springer}  $\tau_K$  of
an atomic level $K$ is defined through the Einstein coefficients $\matha_{KN}$, \frdwa{e1}{e2}, as
\equl{\tau_K = \frac{1}{\sum_{N}\matha_{KN},}
}{lifetime}
where the summation over $N$ includes  all states (channels) to which the state  $K$ can decay. 

All  formulas derived in this work 
were implemented in a suite of programs developed by one of us (AMT):
an ab initio program for the coupled cluster calculations,
the \paldus~program for symbolic manipulations, factorization and automatic generation of orbital expressions suitable
for Open-MP parallel implementation,
and the \wigner~script  for  angular momentum manipulations and
 transformation of the transition moments
 from the point group symmetry basis to the $\ket{ LSJ}$ basis.
 \subsection{Notation}
\begin{center}
\begin{tabular}{ll}
  Symbol & Meaning\\
  \hline
  $\TT(N - M)$ & reduced transition moment from state $N$ to $M$\\
  $\matha(N - M)$ &transition probability from state $N$ to $M$\\
  $\mathdm = (d_{-1}, d_0, d_1)$& dipole moment operator\\
  $\mathqm= (Q_{-2}, Q_{-1}, Q_0, Q_1, Q_2)$& quadrupole moment operator\\
  XCCSD(G)& This work, CCSD approximation, Gaussian basis set\\
  XCC3(G)& This work, CC3 approximation, Gaussian basis set\\ %
  XCCSD(S)&This work, CCSD approximation, Slater basis set\\
  XCC3(S)&This work, CC3 approximation, Slater basis set\\
  \hline
\end{tabular}
\end{center}

\subsection{Basis sets}
In this work we study  several low-lying states of the Ca, Sr, and Ba atoms.
The results were
obtained with  two types of basis sets:
 Gaussian-type orbitals (GTO)\cite{boys1950electronic, boys1956automatic} and
 Slater-type orbitals (STO).\cite{slater1930atomic, zener1930analytic}
 STO basis sets are usually significantly smaller than
 compared with GTO basis sets of a comparable
 quality. Therefore, there is a strong reason to use
 them in the computationally demanding coupled cluster theory.
 STOs  used in this work were constructed according to
 the correlation-consistency principle.\cite{dunning1989gaussian}
For the Ca, Sr, and Ba atoms we used the STO basis sets  specifically designed for the calculations
with the effective core potentials.\cite{lesiuk2017combining}
The following Gaussian basis sets were used:
for Ca the ECP10MDF pseudopotential\cite{lim2006relativistic} together with the $[12s12p6d]$ basis augmented with  a set of $[1s1p1d]$
diffuse functions, 
for Sr  $[8s8p5d4f1g]$  augmented
with a set of $[1s1p1d1f3g]$ diffuse functions\cite{skomorowski2012rovibrational} and the ECP28MDF
pseudopotential,\cite{feller1996role, schuchardt2007basis,  lim2006relativistic}
for Ba the ECP46MDF pseudopotential\cite{lim2006relativistic} together with the $[9s9p6d4f2g]$ Gaussian basis set.\cite{lim2006relativistic}
To assess the quality of the basis sets, 
in \frttt{ca-en}{ba-en} we present the excitation energies obtained with the EOM-CCSD and EOM-CC3 codes and
compare them with the experimental results. In the case of the triplet states
we used the ``non-relativistic''  experimental  values deduced from the Land\'{e} rule.
 \begin{center}
  \begin{table}[ht!]
 \caption{Excitation energies of the calcium atom in cm$^{-1}$.}\label{ca-en}
 \begin{center}
\begin{tabular}{llllll@{}}
State & EXP\cite{sugar1979energy, miyabe2006determination}&CCSD(G)&CCSD(S)&CC3(G)&CC3(S)\\
  \hline
  $\tP^\circ$& 15263.1& 15098.7& 15173.2& 15063.5&15195.3 \\
  $\tD$& 20356.6& \textemdash$^{\mt{a}}$& 20856.1& \textemdash$^{\mt{a}}$&21299.6 \\
  $\sD$& 21849.6& \textemdash$^{\mt{a}}$& 22878.6& \textemdash$^{\mt{a}}$& 22859.0\\
  $\sP^\circ$& 23652.3& 24724.4& 24845.8& 24080.5&23879.6 \\
  $\tS$& 31539.5& 31518.3& 31828.7& 31157.3&31545.5 \\
  $\sS$& 33317.3& 33566.5& 33890.9&32983.0 &\textemdash$^{\mt{a}}$
\end{tabular}
\\ \vspace{0.5cm}
\footnotesize{$^{\mt{a}}$ Iterataions did not converge to the desired state. Explained in the text.}
\end{center}
\end{table}
 \end{center}
    \begin{center}
  \begin{table}[ht!]
 \caption{Excitation energies of the strontium atom in cm$^{-1}$.}\label{sr-en}
 \begin{center}
\begin{tabular}{llllll}
State & EXP\cite{sansonetti2010wavelengths}&CCSD(G)&CCSD(S)&CC3(G)&CC3(S)\\
  \hline
  $\tP^\circ$& 14702.9&14575.6&14546.3&14570.8&14597.2\\
  $\tD$& 18253.8&18414.5&18155.0&18668.8&18393.7\\
  $\sD$& 20149.7&20814.1&20584.7&20650.3&20411.1\\
  $\sP^\circ$&21698.5&22632.7&22701.9&21764.3&21797.5\\
  $\tS$& 29038.8&29137.0&29189.7&28885.3&28939.3\\
  $\sS$& 30591.8&31014.0&31063.1&30464.4&30508.6\\
\end{tabular}
\end{center}
  \end{table}
  \end{center}

    \begin{center}
  \begin{table}[ht!]
 \caption{Excitation energies of the barium atom in cm$^{-1}$.}\label{ba-en}
 \begin{center}
   \begin{tabular}{l d{5.1}d{5.1}d{5.1}d{5.1}d{5.1}}
     \multicolumn{1}{c}{State} &
     \multicolumn{1}{c}{EXP\cite{post1985odd}}&
     \multicolumn{1}{c}{CCSD(G)}&
     \multicolumn{1}{c}{CCSD(S)}&
     \multicolumn{1}{c}{CC3(G)}&
     \multicolumn{1}{c}{CC3(S)}\\
\hline
  $\tD$&9357.8&9270.9&8923.7&9581.6&9178.1\\
  $\sD$& 11395.4&12063.6&11653.5&11869.7&11391.4\\
  $\tP^\circ$& 13085.5&12970.5&12823.6&13069.8&12925.9\\
  $\sP^\circ$&18060.3&19569.0&19527.3&18372.2&18284.6\\
  $\tS$& 26160.3&26136.6&26269.3&24275.7&26141.9\\
  $\sS$& 26757.3&27760.6&27971.9&25826.2&25213.0\\
\end{tabular}
\end{center}
  \end{table}
  \end{center}
It can be seen from \frttt{ca-en}{ba-en} that  our results are in a good agreement with the experimental data with absoulute average deviation (AAD) of about 2\%.
For the Ca atom, most of the states are in a perfect agreement with
the experiment (average error $0.3\%$ for Slater/EOM-CC3 case). However,  there are significant discrepancies for the
$\sD$ and $\tD$ states. Within the Gaussian basis set both EOM-CCSD and EOM-CC3 iterations did not converge to the desired states, and within
the Slater basis set the errors are around $4\%$. An earlier analysis of this problem\cite{lesiuk2017combining} revealed that 
 this is an inherent problem of the pseudopotentials used in the
 calculations. The authors of Ref.~\citenum{lesiuk2017combining}
 noted that this artifact was also observed in the original paper of Lim.\cite{lim2006relativistic}
 
        For the Sr atom we observe AAD from the experiment around of $2\%$ for the CCSD case.
        The best agreement is found in the case of the Slater basis set and the  EOM-CC3 method, where AAD  is only $0.6\%$.
For the Ba atom, the agreement is slightly worse than in the Sr case, with
the average error of  $0.9\%$ in the case of the Slater basis set and the EOM-CC3 method (we assumed that the $\sS$ did not correctly converge, as it gave a $5\%$ error).

We also analyzed the overall importance of triple excitations. Both in Slater and Gaussian basis sets,
AAD from the experiment is lower with the inclusion of the triples excitations. Using the
Slater basis set in the CCSD case we did not observe a significant improvement.
In the case of $\sD$ and $\tD$ states of the Ca atom, the Slater basis set
allowed for the correct convergence of the EOM-CC method.
In the computation of  lifetimes, whenever an energy level for a specific $J$ was required (that is, in all
 cases where the energy of  triplet state was required explicitly), we used the experimental energies.  

 \subsection{The $5s5p\tPj^\circ$ state of the Sr atom}
 In what follows we denote the wave functions by the irreducible representations of the $D_{2h}$ point group 
  $\Gamma = \{\Ag, \Bjg, \Bdg, \Btg, \Au, \Bju, \Bdu, \Btu\}$.
The lifetime of the $5s5p\tPj^\circ$ state is computed as
\equ{\tau_{[\prescript{3}{}{\mbox{\tiny{P}}}_1^\circ]}  = \frac{1}{\matha(\tPj^\circ -\sSz )}.
}
The transition moment from the $\tPj^\circ$ state is expressed in the point group symmetry basis as
\equa{
  \TT(\tPj^{\circ} - \sSz) &= 
  \sqrt{3}\braket{ \sSz^0|z|\sPj^1 }\frac{\braket{ \sPj^1|H_{\rm{SO}}|\tPj^1 }}{E_{\tPjt}-E_{\sPjt}} =\nonumber\\
  &=\sqrt{3}\braket{\sAg|z|\sBju}
  \frac12\frac{(\braket{\sBju|V^x|\tBdu} + \braket{\sBju|V^y|\tBtu})}{E_{\tPjt}-E_{\sPjt}}.
}
A comparison of our results with the existing theoretical and experimental data
is presented in \Frt{sr2a}. 
Skomorowski et al.\cite{skomorowski2012rovibrational} obtained $\tau=21.40$ $\mu$s
using TD-CC3 method together
with the multireference CI for the spin-orbit coupling matrix elements.
Porsev et al.\cite{porsev2001many} obtained $\tau=19$ $\mu$s with the use of the CI+MBPT method. 
We observe that both in the Gaussian and Slater basis sets the inclusion of the triple excitations results in a longer lifetime. The use of the Slater basis set results in a shorter lifetime in the XCCSD and XCC3 cases.
Our computed lifetime $\tau=24.6$ $\mu$s is in a perfect
agreement with the  value obtained by  Santra et al.\cite{santra2004properties} 
($\tau = 24.4$  $\mu$s).
The experimental work from 2006 of Zelevinsky et al.\cite{zelevinsky2006narrow} suggest a lower value of $\tau=21.5$ $\mu$s.

\begin{table}[!htb]
    \caption{Lifetime $\tau$ in $\mu$s of the $5s5p\tPj^{\circ}$ state of the Sr atom.}\label{sr2a}
      \centering
          \begin{tabular}{l d{2.2}cd{1.1}}
   \text{Reference} &     \multicolumn{3}{l}{$\tau$ [$\mu$s]} \\
\hline
\multicolumn{4}{c}{Theory}\\
\hline
                 This work XCCSD(G)    & 23.67&&\\
                 This work XCC3(G)    & 25.00&&\\
                 This work XCCSD(S)    & 23.24&&\\
                 This work XCC3(S)    & 24.60&&\\
                 Skomorowski et al.\cite{skomorowski2012rovibrational} & 21.40&&\\
                 Santra et al.\cite{santra2004properties} & 24.4&&\\
                 Porsev et al.\cite{porsev2001many} & 19.0&&\\

\hline
\multicolumn{4}{c}{Experiment}\\
\hline
                 Zelevinsky et al.\cite{zelevinsky2006narrow} & 21.5 &$\pm$&0.2\\
\end{tabular}
    \end{table}%

\subsection{The $6s5d \tDd$ state of the Ba atom}
We calculated  the lifetime of the 
 $6s5d \tDd$ state. The following expression was derived
using the \wigner~code
\equa{\TT(\tDd, \sSz) & = \braket{6s5d \tDd||\mathq ||6s^2 \sSz} = \sqrt{5} \braket{6s5d \tDd^0|Q_0| 6s^2 \sSz^0}\\
 & =\sqrt{5}  \frac{\braket{6s5d \tDd^0|H_{\rm{SO}}| 6s5d \sDd^0}}{E_{\tDdt^0} - E_{\sDdt^0}}\braket{6s5d \sDd^0|Q_0| 6s^2 \sSz^0}\nonumber\\
  &= \sqrt{5}\frac{\left( -\frac12\braket{\sAg|V^x|\tBtg} + \frac12\braket{\sAg|V^y|\tBdg}\right)}{E_{\tDdt^0} - E_{\sDdt^0}}\nonumber\\
  &\times\sqrt{\frac32} \braket{\sAg|Q_{zz}|\sAg}. \nonumber
}
 \equ{Q_0 = \sqrt{\frac32}Q_{zz}
 }
 In \Frt{trypletba} we present  comparison of our results with the available theoretical data, as no experimental results
 have been reported
 thus far.
 \begin{table}[!ht]
  \begin{center}
    \caption{Lifetimes for the $5s6d\tDd$ [$s$] state for barium atom. (T/E, L/V) denote Theoretical/Experimental energy and
    Length/Velocity representation.}\label{trypletba}
  \begin{tabular}{l l d{4.1}}
    Reference &Method& \multicolumn{1}{c}{$\tau$[s]}\\
                  \hline
                  This work&XCC3(G) & 20.0\\
    \multirow{8}{*}{Migdalek et. al.\cite{migdalek1990multiconfiguration}
         $\begin{dcases*} \\ \\ \\ \\ \\ \\ \\  \end{dcases*}$}  & MCDF-I (T, L) & 418.3 \\
&  MCDF-I (T, V) & 3404.2\\
 &MCDF-I (E, L) & 582.6\\
 &MCDF-I (E, V) & 4153.0\\
 &MCDF-II (T, L) & 43.6\\
 &MCDF-II (T, V) & 50.6\\
 &MCDF-II (E, L) & 59.4\\
 &MCDF-II (E, V) & 60.9\\[0.2ex]
 Trefftz et. al. \cite{trefftz1974mutual}&MCHF& 20.0\\
  \end{tabular}
  \end{center}
\end{table}
Unfortunately, there are very few theoretical results available for this state in the literature, either.
Migdalek et al.\cite{migdalek1990multiconfiguration} employed a relativistic multiconfigurational Dirac-Fock method (MCDF) and
performed two types of calculations. In MCDF-I the relativistic counterparts of only $6s5d$ and $5d^2$ configurations are included, and in MCDF-II
the $6p^2$ configuration is additionally included. The results for the $5s6d\tDd$ Ba lifetime are given both in the length and velocity
representations. It is clear from \Frt{trypletba}
that a huge variation of the results was observed for MCDF-I depending on the choice of the gauge. As the difference
between the length and velocity gauges is frequently used to verify the quality of a method, 
the authors suggest that the MCDF-II method works better in
this case. 
We also compare our results with the computations of Trefftz,\cite{trefftz1974mutual} where multi-configurational Hartree-Fock (MCHF) wave functions
were used in the configuration interaction method including the spin-orbit coupling.
The authors obtained $\tau = 20.0$ s which is in a perfect agreement with our computed value of $\tau=20.0$ s.

\subsection{The $3d4s\sDd$ state of the Ca atom and the $5s4d\sDd$ state of the Sr atom}
The lifetimes of the $3d4s\sDd$ state of calcium and the $5s4d\sDd$  state of strontium are defined as 
\equ{ \tau_{[\prescript{1}{}{\mbox{\tiny{D}}}_2]} =  \frac{1}{\matha(\sDd, \sSz) + \matha(\sDd, \tPd) + \matha(\sDd, \tPj)}.
}
and are  especially interesting  as
 three different transitions contribute to it.
  The expressions for the quadrupole transition  
$\TT(\sDd, \sSz)$ and two spin-forbidden  transitions, $\TT(\sDd, \tPd)$
and $\TT(\sDd, \tPj)$, are 
\equal{
  &\TT(\sDd, \sSz) &&= \braket{ \sDd||\mathqm|| \sSz}  = \sqrt{5}\sqrt{\frac32} \braket{ \sDd^0|Q_{zz}| \sSz^0} \\
&  &&=\sqrt{5}\sqrt{\frac32} \braket{\sAg|Q_{zz}|\sAg},\nonumber\\
  &\TT(\sDd, \tPd) && = 
    -\sqrt{10}\frac{\braket{ \sDd^{\sm1}|H_{\rm{SO}}|\tDd^{\sm1}}}{E_{\sDdt}-E_{\tDdt}} \braket{ \tDd^{\sm1}|d_{\sm1}| \tPd^0}\nonumber\\
&   &&= -\sqrt{10}\frac{\left(-\frac12\braket{\sAg|V^x|\tBtg}+\frac12\braket{\sAg|V^y|\tBdg}\right)}{E_{\sDdt}-E_{\tDdt}} \nonumber\\
&  &&\times \left(\frac14\braket{\tBju|x|\tBdg} - \frac14\braket{\tBju|y|\tBtg}\right),\nonumber\\
&\TT(\sDd, \tPj)&&=
-\sqrt{\frac{15}{2}}
    \frac{\braket{ \sDd^{\sm1}|H_{\rm{SO}}|\tDd^{\sm1}}}{E_{\sDdt}-E_{\tDdt}} \braket{ \tDd^{\sm1}|d_{\sm1}| \tPj^0}\nonumber\\
    &   &&-\sqrt{\frac{15}{2}} \frac{\braket{ \tPj^{0}|H_{\rm{SO}}|\sPj^{0}}}{E_{\sPjt}-E_{\tPjt}} \braket{ \sPj^{0}|d_0| \sDd^0} \nonumber\\
&  &&=-\sqrt{\frac{15}{2}}
      \frac{ \left(-\frac12\braket{\sAg|V^x|\tBtg}+\frac12\braket{\sAg|V^y|\tBdg}\right)}{E_{\sDdt}-E_{\tDdt}}\nonumber\\
&      &&\times\big(\frac12\braket{\tBdu|z|\tBtg}+\frac12\braket{\tBdg|z|\tBtu}\big)
     \nonumber\\
&     &&-\sqrt{\frac{15}{2}}\frac{\left(-\frac12\braket{\tBju|V^x|\tBdu}+\frac12\braket{\tBju|V^y|\tBtu}\right)}{E_{\sPjt}-E_{\tPjt}}
     \braket{ \sAg|z|\tBju}.\nonumber
}{wig1}

\begin{table}[!htb]
  \caption{Lifetime $\tau$ in ms of the $3d4s\sDd$ state of the Ca atom.}\label{ca3a}
      \centering
          \begin{tabular}{l d{1.2}cd{1.3}}
    \text{Reference} &     \multicolumn{3}{l}{$\tau$ [ms]} \\
\hline
\multicolumn{4}{c}{Theory}\\
\hline

                This work XCC3(S)    & 3.5 & &\\
                Bauschlicher et al.\cite{bauschlicher1985radiative} & 3.05 & &\\
                 Bauschlicher et al.\cite{bauschlicher1985radiative} & 2.76 & &\\
\hline
\multicolumn{4}{c}{Experiment}\\
\hline
                Beverini et al.\cite{beverini2003measurementof} & 2.3 &$\pm$ &0.5\\
                Drozdowski et al.\cite{drozdowski1993lifetimes} & 1.5 &$\pm$ &0.4\\
                Pasternack et al.\cite{pasternack1980experimental} & 2.3 &$\pm$ &0.5\\

\end{tabular}
\end{table}%

\begin{table}[!htb]
          \caption{Lifetime $\tau$ in ms of the $5s5p\sDd$ state of the Sr atom.}\label{sr3a}
      \centering
          \begin{tabular}{l d{1.2}cd{1.3}}
   \text{Reference} &     \multicolumn{3}{l}{$\tau$ [ms]} \\
\hline
\multicolumn{4}{c}{Theory}\\
\hline
                This work XCCSD(G)    & 0.43&&\\
                This work XCC3(G)    & 0.52&&\\
                This work XCCSD(S)    & 0.36&&\\
This work XCC3(S)    & 0.34&&\\
Skomorowski et al.\cite{skomorowski2012rovibrational} & 0.23&&\\
Bauschlicher et al.\cite{bauschlicher1985radiative} & 0.49&&\\
\hline
\multicolumn{4}{c}{Experiment}\\
\hline
 Courtillot et al.\cite{courtillot2005accurate} & 0.30&&\\
  Husain and Roberts\cite{husain1988radiative} & 0.41&$\pm$&0.01\\
\end{tabular}
\end{table}%
In \Frt{ca3a} we present the lifetime of the $3d4s\sDd$ state of the Ca atom  computed with
XCC method and compare it with the available theoretical and experimental data.
Lifetimes computed with the XCCSD(G), XCCSD(S), XCC3(G) methods are not available due to the divergence of the CC procedure.\cite{lesiuk2017combining}
It can be seen from  \fr{wig1}  that in order to compute all the components
that contribute to $\tau_{[\prescript{1}{}{\mbox{\tiny{D}}}_2]}$
we need the EOM-CC procedure to converge to the $\sD, \tD, \sP$ and $\tP$ states.
As mentioned in the discussion of \Frt{ca-en} in some cases we did not achieve
convergence to the proper states so we were unable to use these states in the transition moment calculations.
The only reliable result was obtained with the XCC3(S) method.
All theoretical methods included in \Frt{ca3a} predict a longer lifetime than
the lifetime measured in experiments. The fact that the XCC3(S) lifetime 
might deviate from the experiment is already indicated by the poor quality of the computed excitation energies for
the $D$ states, which deviates from the reference by roughly 1000 $cm^{-1}$.

For the Sr atom, \Frt{sr3a},  we see that the use of the Slater
basis set gives shorter lifetimes than the use of the Gaussian basis set. Inclusion of triple
excitations with  the Gaussian basis set gives a longer lifetime than for CCSD case. 
In the Slater basis set the trend is opposite. As the EOM-CC3(S) states are of the best
quality (see \frs{sr-en}), we compare the XCC3(S) value with the experiment.
As shown in \Frt{sr3a}, the existing experimental and theoretical results
are scattered on the interval from 0.30 to 0.49 ms.
Our computed lifetime $0.34$ ms lies within this range and simultaneously 
 is close to the
most recent (2005) experimental result $\tau=0.30$ ms of Courtillot et al.\cite{courtillot2005accurate}

\section{Summary and Conclusions}

The extension of the expectation value coupled cluster method (XCC) to the computation of
spin-orbit coupling matrix elements is reported.
The spin-orbit interaction was treated perturbatively by computing the matrix element of the SO part of the pseudopotential.
This approach allowed us to test the performance of our
method for medium and heavy atoms where the SO interaction contributes significantly.
The methodology presented here can be extended to CC models other than CCSD and CC3. 
Our final result, \fr{glowne}, is presented in a commutator form, and can be approximated systematically to include higher excitations. 

To apply the formula for the transition matrix element, \fr{glowne}, we perform its commutator expansion and retain
the contributions according to their leading MBPT order. The amplitudes that enter the formula for  the transition moment come from 
the CCSD or CC3 calculations and the Jacobian eigenvectors are computed with the EOM-CCSD or EOM-CC3 methods.
Our conclusion is that the third order of MBPT is sufficient to obtain converged results.

The accuracy of our method was tested on selected states of the Ca, Sr and Ba atoms.
We compared our results with other theoretical data available in the literature and discussed possible reasons for the observed differences. 
The most important effect was that in some cases the use of the Slater basis set allowed for the proper convergence to the desired states. From the computation of the excitation energies we deduced that the inclusion of triples on average improved the results.
For all of the computed excitation energies the CC3 approximation lowered the average deviation from the experimental data.
Our  conclusion is that
the best choice is to use the Slater basis set and the CC3 level of approximation.
There is a room to extend the XCC theory for the calculation of the magnetic moments, nonadiabatic couplings and open-shell systems.
Currently we work  on combining  the explicitly correlated $\rm{F}12$ approach with the XCC theory.
\section{Acknowledgment}
We would like to thank Dr. M. Modrzejewski for fruitful discussions and detailed reading of the manuscript.
 This research was supported by the National Science Center (NCN) under Grant No. 2017/25/B/ST4/02698.


%

\end{document}